\documentclass{jpsj-suppl}
\usepackage{times}
\usepackage{color}

\title{Charge-density-wave formation in the Edwards fermion-boson model
       at one-third band filling}

\author{
Satoshi Ejima\thanks{E-mail address: ejima@physik.uni-greifswald.de}
and Holger Fehske\thanks{E-mail address: fehske@physik.uni-greifswald.de} 
}
\inst{
Institut f\"ur Physik, Ernst-Moritz-Arndt-Universit\"at
Greifswald, 17489 Greifswald, Germany
}

\recdate{\today}

\abst{
We examine the ground-state properties of the one-dimensional Edwards 
spinless fermion transport model by means of large-scale density-matrix renormalization-group   
calculations. Determining the single-particle gap  and the Tomonaga-Luttinger 
liquid parameter ($K_\rho$) at zero temperature, we prove the existence of a metal-to-insulator 
quantum phase transition at one-third band filling.  The insulator---established by strong correlation in the background medium---typifies a 
charge density wave (CDW)  that is commensurate with the band filling. $K_\rho=2/9$ is very small at the quantum critical point, 
and becomes $K_\rho^{\rm CDW}=1/9$ in the infinitesimally doped three-period CDW, as predicted by the bosonization approach.
}

\kword{Edwards model, metal-insulator transition, DMRG}

\begin{document}
\maketitle

\section{Introduction}
Strong correlations can affect the transport properties of low-dimensional systems to the point of insulating behavior. 
Prominent examples are  broken symmetry states of quasi one-dimensional (1D) metals, where charge- or spin-density 
waves brought about by electron-phonon or by electron-electron interactions~\cite{Gr94}. These interactions 
can be parametrized by bosonic degrees of freedom, with the result that the fermionic charge carrier becomes ``dressed'' by a boson
cloud that lives in the particle's immediate vicinity and takes an active part in its transport~\cite{Be09}.  
A paradigmatic model describing quantum transport in such a ``background medium'' 
is the  Edwards fermion-boson model~\cite{Ed06,AEF07}. 
The model exhibits a surprisingly rich phase diagram including metallic repulsive and attractive 
Tomonaga-Luttinger-liquid (TLL) phases, insulating charge-density-wave (CDW) 
states~\cite{WFAE08,EHF09,EF09b,FEWB12}, and even regions where phase
separation appears~\cite{ESBF12}.
 
The part of the Edwards Hamiltonian that accommodates boson-affected transport is 
\begin{eqnarray}
 H_{fb}= -t_b\sum_{\langle i, j \rangle} f_j^{\dagger}f_{i}^{\phantom{\dagger}}
  (b_i^{\dagger}+b_j^{\phantom{\dagger}})\,.
\end{eqnarray}
Every time a spinless fermion hops between nearest-neighbor lattice sites $i$ and $j$ it creates (or absorbs) a local boson 
$b_j^{\dagger}$ ($b_i^{}$). As to  $H_{b}=\omega_0\sum_i b_i^\dagger b_i^{\phantom{\dagger}}$ this enhances (lowers) the energy of the background by $\omega_0$.  Moving in one direction only, the fermion  creates
a string of local bosonic excitations that will finally  immobilize the particle (just as for a hole in a classical N\'{e}el background). 
Because of quantum fluctuations any distortion in the background should be able to relax however. Incorporating this 
effect the entire Edwards model takes the form
\begin{eqnarray}
 H=H_{fb}-\lambda\sum_i(b_i^{\dagger}+b_i^{\phantom{\dagger}})+H_b\,,
\label{model}
\end{eqnarray}
where $\lambda$ is the relaxation rate.  The unitary transformation $b_i\to b_i + \lambda/\omega_0$ replaces the
second term in~(\ref{model}) by  a direct, i.e., boson-unaffected, fermionic hopping 
term $H_f=-t_f\sum_{\langle i, j \rangle} f_j^{\dagger}f_{i}^{}$. In this way the particle
can move freely, but with a renormalized transfer amplitude  $t_f=2\lambda t_b/\omega_0$.
We note that coherent propagation of a fermion is possible even in the limit $\lambda=t_f=0$, 
by means of a six-step vacuum-restoring hopping being related to an effective next-nearest-neighbor transfer.  
This process takes place on a strongly reduced energy scale (with weight $\propto t_b^6/\omega_0^5$), and is 
particularly important in the extreme low-density regime ($n^f\ll 1$), where the Edwards model  mimics the motion 
of a single hole in a quantum antiferromagnet~\cite{EEAF10}. 

At low-to-intermediate particle densities $n^f \leq 0.3$ the 1D Edwards model 
system stays metallic. If here the fermions couple to slow (low-energy) bosons ($\omega_0/t_b \lesssim 1)$, 
the primarily repulsive TLL becomes attractive, and eventually even phase segregation into particle-enriched and particle-depleted regions takes place at small $\lambda$~\cite{ESBF12}. No such particle attraction is observed, however, 
for densities $0.3\lesssim n^f\leq 0.5$.  Perhaps, in this regime, the repulsive TLL might give way to an insulating state with
charge order if the background is ``stiff'', i.e., for small $\lambda/t_b$  and  fast (high-energy) bosons $\omega_0/t_b >1$. 
So far, a correlation induced TLL-CDW metal-insulator transition like that has been proven to exist for the half-filled band case 
($n^f=0.5$)~\cite{WFAE08,EHF09}.  In the limit $\omega_0/t_b\gg 1 \gg \lambda/t_b$ the Edwards model can be approximated 
by an effective  $t$-$V$ model, $H_{tV}=H_f+V\sum_i n_i^f n_{i+1}^f$,  with nearest-neighbor Coulomb interaction 
$V=t_b^2/\omega_0$~\cite{NEF13}. The spinless fermion $t$-$V$ model on his part can be mapped onto the 
exactly solvable $XXZ$-Heisenberg model, which exhibits a Kosterlitz-Thouless~\cite{KT73} (TLL-CDW) quantum 
phase transition at $(V/t_f)_c=2$, i.e., at $(\lambda/t_b)_{tV,c}=0.25$. The critical value is in reasonable agreement with that 
obtained for the half-filled Edwards model in the limit  $\omega_0\to \infty$: $(\lambda/t_b)_c \simeq  0.16$~\cite{EHF09}. 
At lower densities, however, for example  at $n^f=1/3$, a CDW instability occurs in 1D $t$-$V$-type models only 
if (substantially large) longer-ranged Coulomb interactions were included, such as a next-nearest-neighbor term $V_2$~\cite{SW04}.

In order to clarify whether the 1D Edwards model by itself shows a metal-to-insulator transition off half-filling at large $\omega_0$ and 
what is the reason for the absence of phase separation for small $\omega_0$, in this work,  we investigate the model at one-third 
band filling, using the density matrix renormalization group (DMRG) technique~\cite{Wh92} combined with 
the pseudo-site approach~\cite{JW98b,JF07} and a finite-size analysis.
This allows us to determine the ground-state phase diagram of the 1D Edwards model 
in the complete parameter range.

\section{Theoretical approach}
To identify the quantum phase transition between the metallic TLL and insulating CDW  phases we inspect---by means of DMRG---the 
behavior of  the local fermion/boson densities $n_i^{f/b}$, of the single-particle gap $\Delta_c$, and  of the the TLL parameter $K_\rho$. 
In doing so, we take into account up to four pseudo-sites, and ensure that the local boson density of the last 
pseudo-site is always less than $10^{-7}$ for all real lattice sites $i$. We furthermore keep up to $m=1200$
density-matrix eigenstates in the renormalization process to guarantee a discarded weight smaller than $10^{-8}$.

For a finite system with $L$ sites  the single-particle charge gap is given by
\begin{eqnarray}
 \Delta_c(L)=E(N+1)+E(N-1)-2E(N),
\end{eqnarray}
where $E(N)$ and $E(N\pm1)$ are the ground-state energies 
in the $N$- and ($N\pm1$)-particle sectors, respectively. In the CDW state $\Delta_c$ is finite,  
but will decrease exponentially across the MI transition point if the transition is of Kosterlitz-Thouless type 
as for the $t$-$V$ model. This hampers an accurate determination of the TLL-CDW transition line. 

In this respect the TLL parameter $K_\rho$ is more promising. Here bosonization field theory  
predicts how $K_\rho$ should behave at a quantum critical point.
In order to  determine $K_\rho$  accurately by DMRG,
we first have to calculate the static (charge) structure factor  
\begin{eqnarray}
 S_c(q)=\frac{1}{L}\sum_{j,l}e^{{\rm i}q(j-l)}
          \langle 
           (f_j^\dagger f_j^{\phantom{\dagger}}-n)
           (f_l^\dagger f_l^{\phantom{\dagger}}-n)
          \rangle\,,  
\end{eqnarray}
where the momenta $q=2\pi m/L$ with integers $0<m<L$~\cite{EGN05}.
The TLL parameter $K_\rho$ is proportional to the slope of $S_c(q)$ in the long-wavelength limit $q\to0^+$:
\begin{eqnarray}
 K_\rho=\pi\lim_{q\to 0}\frac{S_c(q)}{q}\,.
\end{eqnarray}
For a spinless-fermion system with one-third band filling, the TLL parameter  should be $K_\rho^\ast=2/9$ at the metal-insulator transition point.  For an infinitesimally doped three-period CDW insulator, on the other hand, 
bosonization theory yields  $K_\rho^{\rm CDW}=1/9$~\cite{Sc94,Gi03}.

\section{Numerical results}
\begin{figure}[tb]
\begin{center}
\includegraphics[clip,width=0.8\columnwidth]{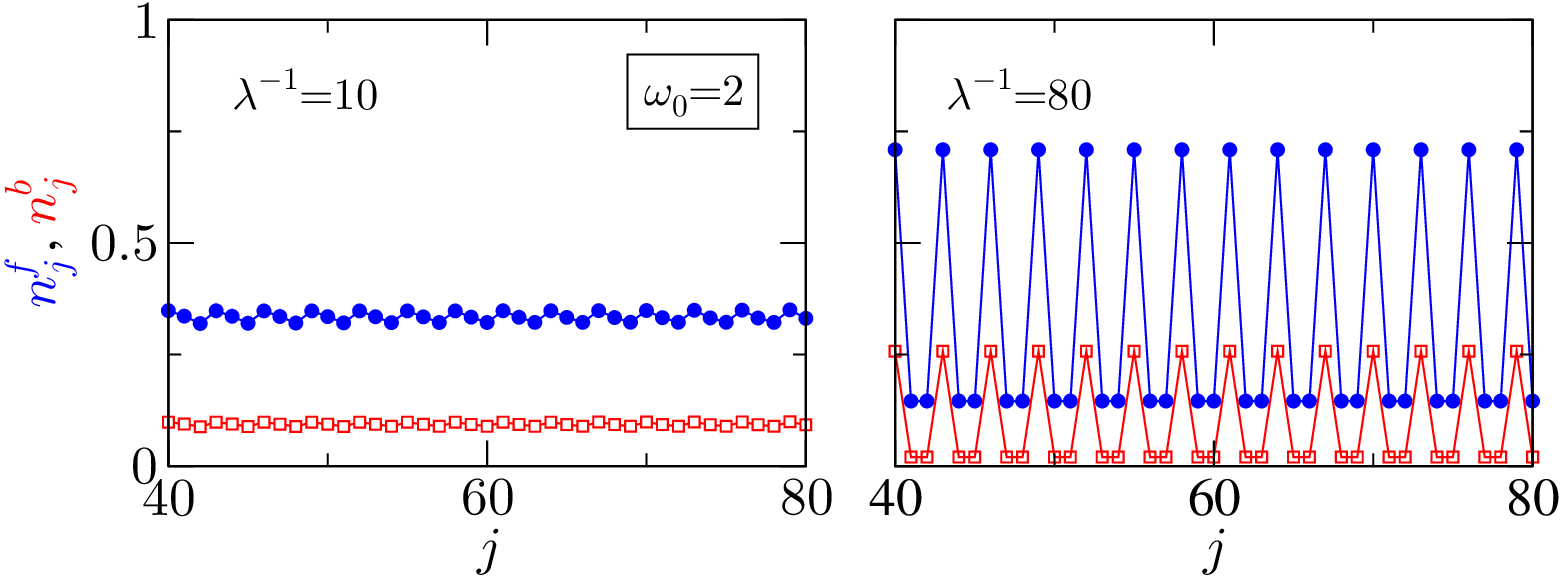}
\end{center}
\caption{(Color online) Local fermion ($n_j^f$ -- filled blue  circles)
 and boson ($n_j^b$ -- open red squares) densities in the  central part of an 
 Edwards model chain with $L=120$ sites and OBC. 
 DMRG data shown in the left-hand (right-hand) panel indicate a homogeneous 
 TLL (CDW) state for $n^f=1/3$ and $\lambda^{-1}=10$ ($\lambda^{-1}=80$), where $\omega_0=2$.
 In what follows all energies are measured in units of $t_b$.}
\label{fig1}
\end{figure}

First evidence for the formation of  a CDW state in the one-third filled Edwards model comes from the spatial variation  
of the local densities of fermions $n_i^f\equiv\langle f_i^\dagger f_i^{\phantom{\dagger}}\rangle$ 
and bosons $n_i^b\equiv\langle b_i^\dagger b_i^{\phantom{\dagger}}\rangle$. Fixing $\omega_0=2$, we 
find a modulation of the particle density commensurate with the band filling factor 1/3 for very small $\lambda=0.0125$ 
(see Fig.~\ref{fig1}, right panel). Thereby, working with open boundary conditions (OBC), one of the three 
degenerate ground states with charge pattern (... 100100100 ...), (... 010010010 ...),
or (... 001001001 ...) is picked up by initializing the DMRG algorithm. As a result the CDW 
becomes visible in the local density. Note that also in the metallic state, which is realized already for
$\lambda$'s as small as 0.1 (cf. Fig.~\ref{fig1}, left panel), a charge modulation is observed. Those, however, can be attributed 
to Friedel oscillations, which are caused by the OBC and will decay algebraically in the central part of the chain 
as $L$ increases. Thus, for $\omega_0=2$, a metal-to-insulator transition is expected to occur in between $10<\lambda^{-1}<80$.

\begin{figure}[t]
 \begin{tabular}{cc}
  \begin{minipage}{0.5\hsize}
   \begin{center}
    \includegraphics[clip,width=0.72\columnwidth]{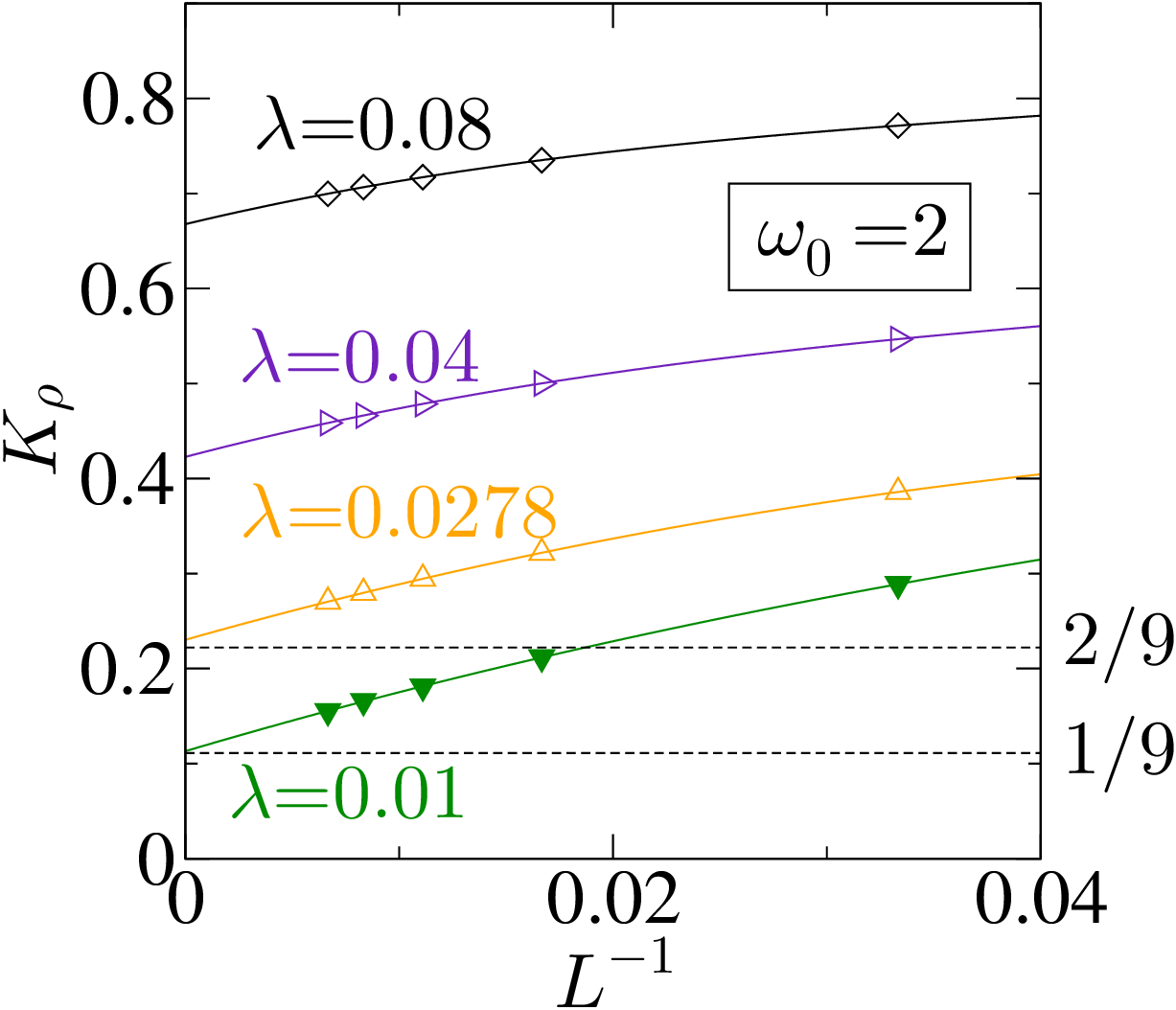}
   \end{center}
  \end{minipage}
\begin{minipage}{0.5\hsize}
 \begin{center}
    \includegraphics[clip,width=0.95\columnwidth]{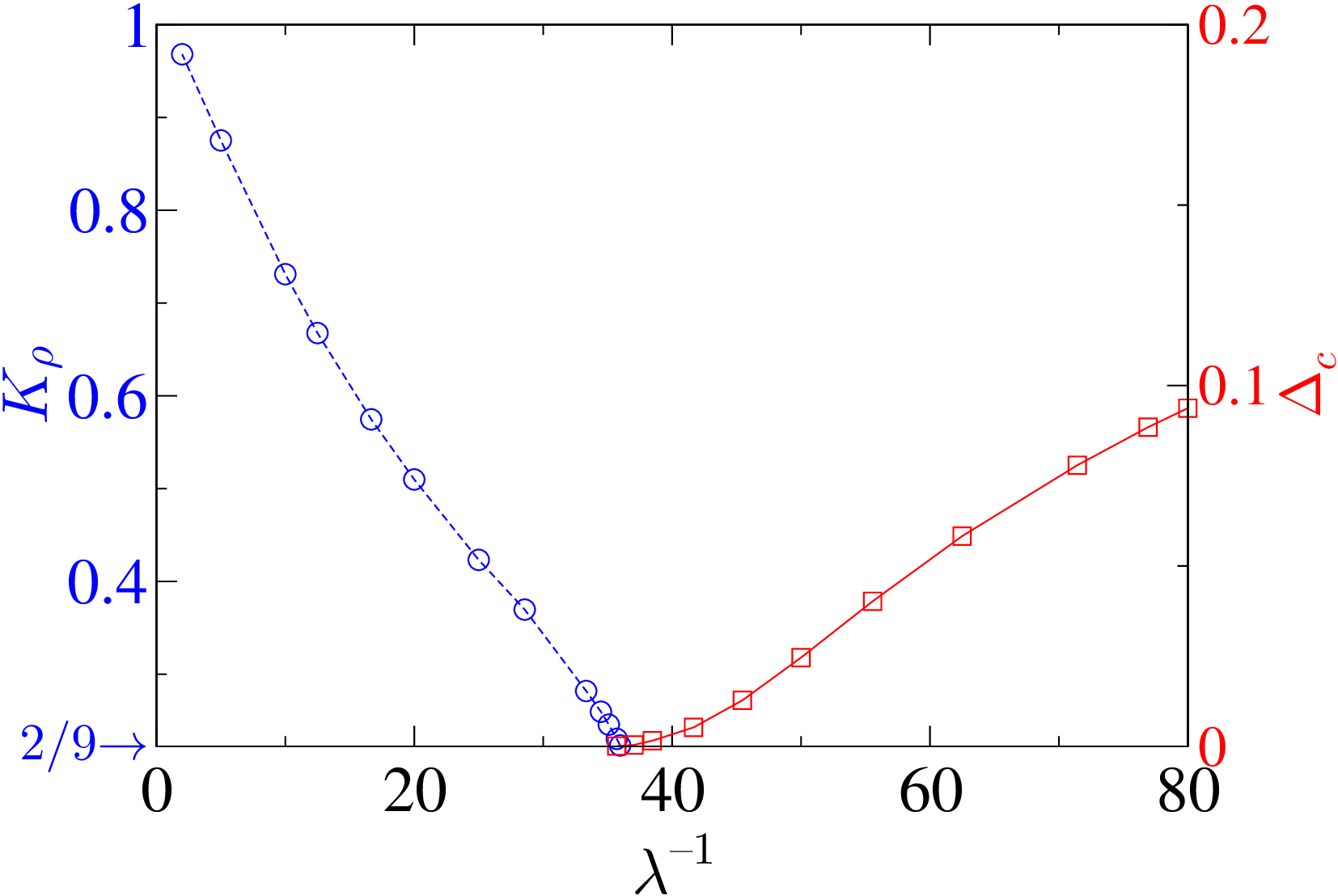}
 \end{center}
\end{minipage}
 \end{tabular}
  \caption{(Color online) Left panel: $K_\rho(L)$ in the one-third filled Edwards model
  as a function of the inverse system size for various values of 
 $\lambda$ at $\omega_0=2$ (open symbols).  The finite-size interpolated DMRG data at the metal-insulator transition point and 
 for the infinitesimally doped CDW insulator [$n^f=1/3-1/L$ (filled symbols)] are in perfect 
 agreement with the bosonization results $K_\rho^\ast=2/9$ and
 $K_\rho^{\rm CDW}=1/9$, respectively.  
 Right panel: $L\to\infty$ extrapolated $K_\rho$ (circles) and  
 $\Delta_{c}$ (squares), as functions of $\lambda^{-1}$ for $\omega_0=2$, indicate a 
 TLL-CDW transition at $\lambda^{-1}\sim 36$.}
 \label{fig2}
\end{figure}

To localize the point where---at given $\omega_0$ and $\lambda$---the quantum phase  transition takes place, we first compute 
the single-particle gap $\Delta_c$ and TLL charge exponent $K_\rho$ for finite chains with up to $L=150$ 
sites and OBC. Then we perform a finite-size scaling as illustrated for $K_\rho$ by  Fig.~\ref{fig2}, left panel.
Here open symbols give $K_\rho$ as a function of the inverse system size $L^{-1}$. 
The DMRG data can be extrapolated to the thermodynamic limit by third-order polynomial functions. 
Decreasing $\lambda$ at fixed $\omega_0=2$ the values of $K_\rho$ 
decreases too and becomes equal to $K_\rho^\ast=2/9$ at the Kosterlitz-Thouless  transition point 
$(\lambda^{-1})_c \sim 36$; see Fig.~\ref{fig2}, right panel. 
For $\lambda^{-1}>36$ the system embodies a $2k_{\rm F}$-CDW insulator with 
finite charge gap $\Delta_c$. Furthermore,  calculating $K_\rho(L)$  for $N=L/3-1$ particles, 
 we can show that the infinitesimally doped CDW insulator has  $K_\rho^{\rm CDW}=1/9$ at $n^f=1/3$. 
 Deep in the CDW phase, $K_\rho$ approaches $1/9$ in the thermodynamic limit 
[cf. the $\lambda=0.01$ data (filled symbols) in the left panel of Fig.~\ref{fig2}].

\begin{figure}[t]
 \begin{tabular}{cc}
  \begin{minipage}{0.5\hsize}
   \begin{center}
    \includegraphics[clip,width=0.95\columnwidth]{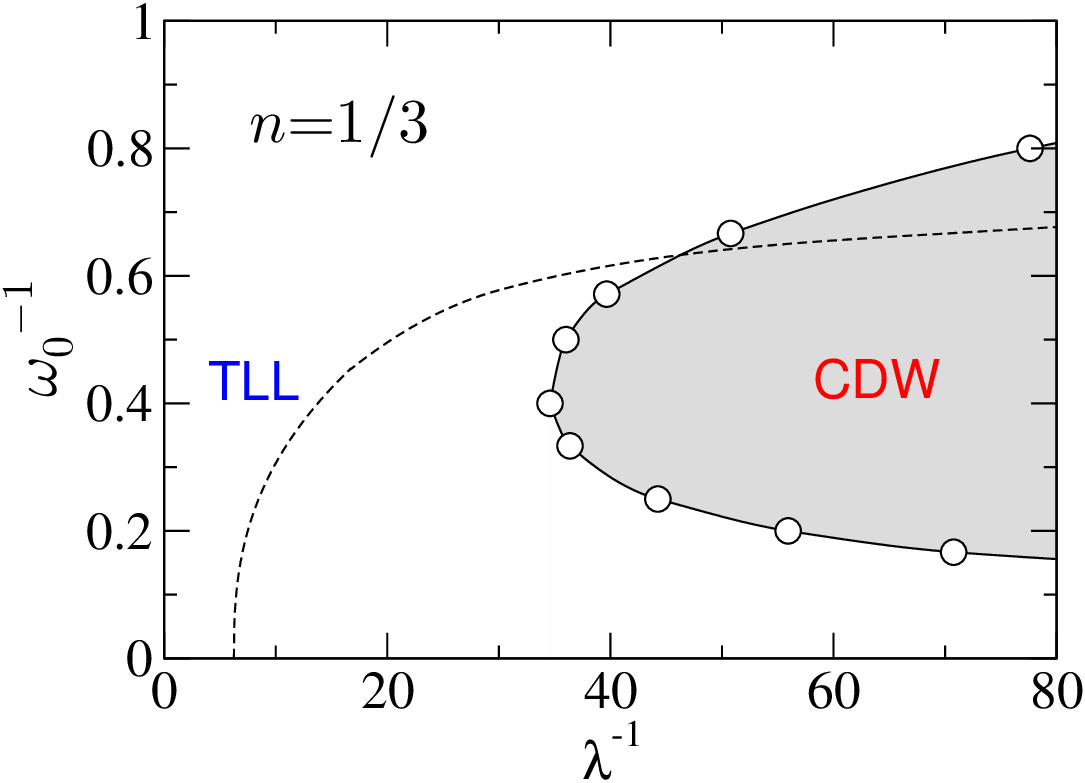}
   \end{center}
  \end{minipage}
\begin{minipage}{0.5\hsize}
 \begin{center}
    \includegraphics[clip,width=0.92\columnwidth]{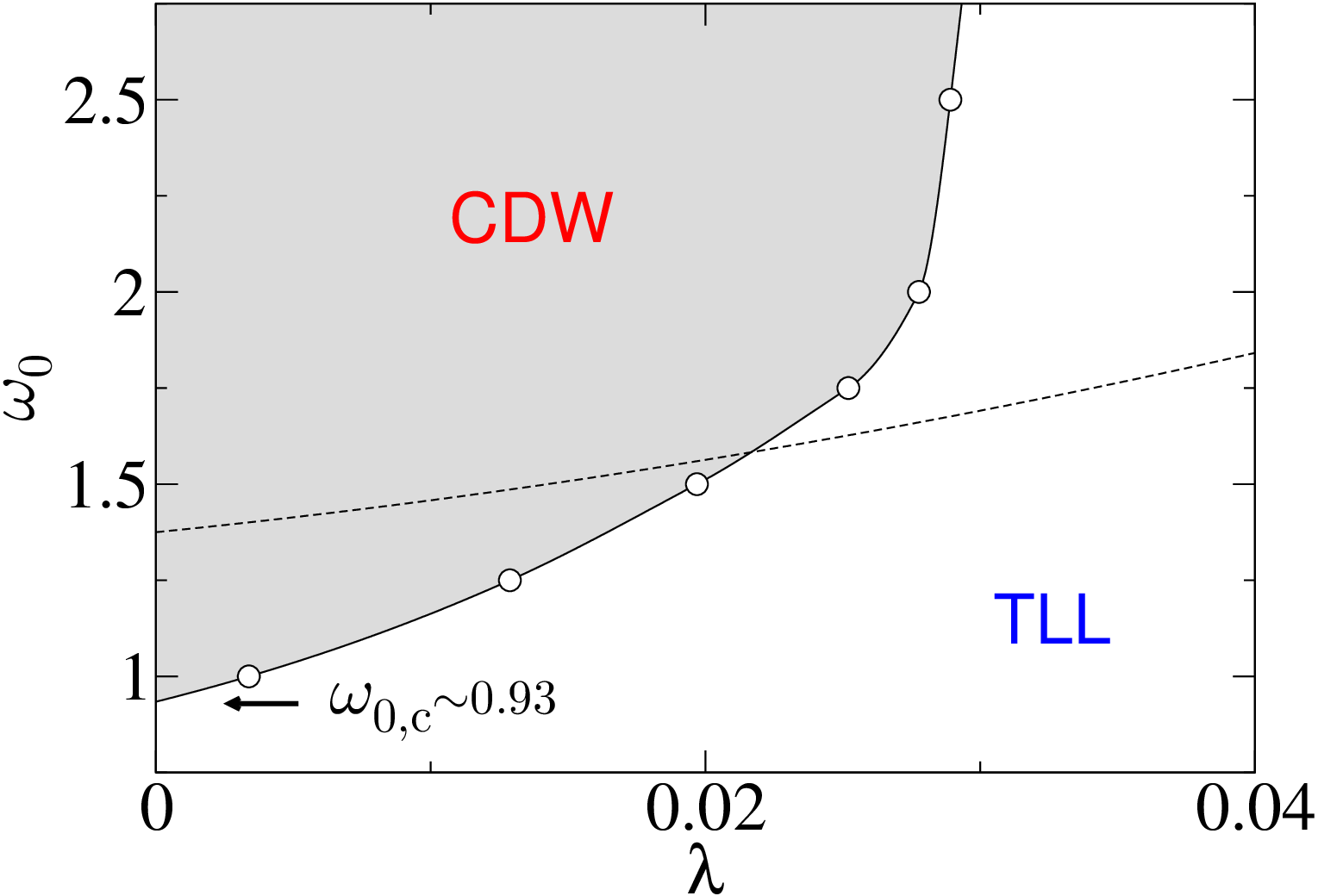}
 \end{center}
\end{minipage}
 \end{tabular}
 \caption{(Color online) DMRG ground-state phase diagram of the 1D Edwards
 model at one-third band filling, showing the stability regions of metallic TLL and insulating CDW phases  
  in the $\lambda^{-1}$-$\omega_0^{-1}$ (left panel) and $\lambda$-$\omega_0$ (right panel) plane.
 The dashed line denotes the MI transition points at half
 band filling from Ref.~\cite{EHF09}.}
 \label{fig3}
\end{figure}

Our final result is the ground-state phase diagram of the 
one-third filled Edwards model shown in Fig.~\ref{fig3}.
The TLL-CDW phase boundary is derived from the 
$L\to\infty$ extrapolated $K_\rho$ values. Within the TLL region 
$2/9<K_\rho<1$. Of course, the TLL appears at large $\lambda$, when any 
distortion of the background medium readily relaxes $(\propto \lambda$), 
or, in the opposite limit of small  $\lambda$, when the rate of the 
bosonic fluctuations ($\propto\omega_0^{-1}$) is sufficiently high. 
Below $\omega_{0,c}\simeq 0.93$ the metallic state is stable $\forall \lambda$,
because the background medium is easily disturbed and therefore 
does not hinder the particle's motion much.  Note that this value is smaller
than the corresponding one for the half-filled band case, where $\omega_{0,c}\simeq 1.38$. 
On the other hand, the $2k_{\rm F}$-CDW phase with $\Delta_c>0$ and long-range order appears, at half-filling, 
for small $\lambda$ and by trend large $\omega_0$ (see dashed lines); $\lambda_c\simeq 0.16$ for $\omega_0\to\infty$~\cite{EHF09}.
Interestingly, for $n^f=1/3$, we observe that the CDW will be suppressed again if the energy of a background distortion 
becomes larger than a certain $\lambda$-dependent  value (see Fig.~\ref{fig3}, left panel). In stark contrast to the half-filled band
case, at $n^f=1/3$, it seems that  the TLL is stable  $\forall \lambda$, 
when $\omega_0\to\infty$.   
This is because in this limit in the corresponding one-third filled  $t$-$V$ model not only a nearest-neighbor Coulomb repulsion  $V$ 
but also a substantial next-nearest-neighbor interaction $V_2$ is needed to drive the TLL-to-CDW transition~\cite{SW04}.  
Again in the limit $\omega_0/t_b\gg 1 \gg \lambda/t_b$, the Edwards model
at one-third filling can be described by the effective $t$-$V$-$V_2$ model
with $V=2t_b^2/3\omega_0$ and $V_2=8t_b^4/3\omega_0^3$, i.e.,
$V_2/t_f=4t_b^3/3\lambda \omega_0^2$, which clearly explains the absence of 
the CDW phase for $\omega_0\gg 1$.

\section{Conclusions}
To summarize, using an unbiased numerical  (density matrix renormalization group) technique, we investigated  
the one-dimensional fermion-boson Edwards model at one-third band filling. We proved that the model 
displays a metal-insulator quantum phase transition induced by correlations in the background medium. 
The metallic phase is a Tomonaga-Luttinger liquid with $2/9<K_\rho<1$. The insulator represents
a $2k_{\rm F}$ charge density wave with  $K_\rho^{\rm CDW}=1/9$ deep inside the long-range ordered
state.  Performing a  careful finite-size scaling analysis, the phase transition point can be precisely determined by $K_\rho$.  
If the background medium is stiff, we can conclude---by analogy with the ground-state 
phase diagram of the one-third filled $t$-$V$-$V_2$ model---that the Edwards model  incorporates the effects 
of both effective nearest-neighbor and next-nearest-neighbor Coulomb interactions 
between the fermionic charge carriers. The effect of the latter one is reduced when the energy of a local distortion 
in the background is very large, which maintains metallic behavior---different from the half-filled band case---even for weak boson relaxation. 

\section*{Acknowledgements}
The authors would like to thank S. Nishimoto for useful discussions.
This work was supported by the Deutsche Forschungsgemeinschaft  through SFB 652, project  B5.

\end{document}